\documentclass[english,a4paper,12pt]{article}
\usepackage[T1]{fontenc}
\usepackage{babel}
\usepackage{graphicx}
\usepackage{epsf}
\usepackage{cite}
\usepackage[dvips]{color}

\begin{document}
\noindent
~ \normalsize \hfill {\bf October 5, 2005}
\begin{center}
{\bf \Large Comments on the PLA article `UCN anomalous losses and the UCN capture
cross section on material defects' [Phys.\ Lett.\ A 335 (2005) 327].}
\end{center}
\normalsize
\begin{center}
M.\,Daum$^{1,a}$, P.\,Geltenbort$^2$, R.\,Henneck$^1$,
K.\,Kirch$^1$  \\
$^1$PSI, Paul Scherrer Institut, CH 5232 Villigen PSI, Switzerland \\
$^2$ILL, Institut Laue-Langevin, 38024 Grenoble Cedex 9, France\\ 
\end{center}

\begin{center}
{\bf \large Abstract}
\end{center}
We comment on the paper 
`UCN anomalous losses and the UCN capture
cross section on material defects' by A.\ Serebrov et al.,
Phys.\ Lett.\ A 335 (2005) 327 - 336. Data presented do not originate
from these authors alone but were taken in collaboration with
several other authors and institutes not mentioned. 

\noindent
{\bf PACS numbers:}\\
28.20.-v \\
{\bf Key words:} \\
Ultracold neutrons, Anomalous losses
\vfill

\noindent
$^a$Corresponding author; e-mail: manfred.daum@psi.ch

\newpage
\noindent
Recently, the article 
`UCN anomalous losses and the UCN capture cross section
on material defects' appeared in Physics Letters A, see Ref.\cite{serebrov}. 
The authors are
A.\ Serebrov, N.\ Romanenko, O.\ Zherebtsov, M.\ Lasakov, A.\ Vasiliev,
A.\ Fomin, P.\ Geltenbort, I.\ Krasnoshekova, A.\ Kharitonov,
V.\ Varlamov from the  Petersburg Nuclear Physis Institute (PNPI), Gatchina, Russia
and Institut Max von Laue-Paul Langevin (ILL), Grenoble, France.
The results presented in 
Figs.\,2 and 4 in this article originate from data taken in joint 
collaboration between PNPI, Paul Scherrer Institut (PSI), ILL, TU Munich,
and University of Sussex, unpublished so far, published in Ref.\cite{serebrov}
by a rather incomplete list of authors and institutes from the mentioned collaboration.

\noindent
{\bf  1) Data in Fig.\,2 of PLA 335 (2005) 327}\\
Part of the data in Fig.\,2, the data for the narrow Be-coated
copper trap at 90\,K, stem from joint experiments of a collaboration
between PNPI Gatchina, 
PSI Villigen, ILL Grenoble and 
Phys.\ Dep.\ TU Muenchen.
The ILL proposals are:\\
(i) ILL proposals 3-14-148 and 3-14-138,\\
''Test for Be coatings for a new Spallation Ultracold Neutron
Source (SUNS) at PSI.''
Authors: A.\ Serebrov (PNPI), M.\ Daum (PSI), R.\ Henneck (PSI),
K.\ Kirch (PSI), P.\ Geltenbort (ILL), K.\ Schreckenbach (TU Munich),
V.\ Varlamov (PNPI), A.\ Kharitonov (PNPI).
 
\noindent
(ii) ILL Proposal 3-14-120,\\
''Precision measurement of the free neutron lifetime using a 
gravitational trap for ultra-cold neutrons (UCN).''
Authors: A.\ Serebrov (PNPI), P.\ Geltenbort (ILL), K.\ Schreckenbach (TU Munich),
J.\ Pendlebury (Sussex), M.\ Daum (PSI), A.\ Pichlmaier (TU Munich),
A.\ Kharitonov (PNPI), V.\ Varlamov (PNPI), R.\ Taldaev (PNPI),
G.\ Shmelev (PNPI), I.\ Krasnochtchekova (PNPI).

Part of the results from these experiments appeared (unrefereed) as a contribution to 
the PSI internal annual report\cite{PSI}.
Here, the authors were
A.\ Kharitonov, I.\ Krasnoschekova, A.\ Serebrov, R.\ Taldaev, V.\ Varlamov, A.\ Vassiljev
(all PNPI), P.\ Geltenbort (ILL), M.\ Daum, R.\ Henneck, K.\ Kirch (all PSI), 
K.\ Schreckenbach (TU Munich).
 
Results from the above mentioned experiments at ILL
from which the data of Fig.\,2 in Ref.\cite{serebrov} were deduced,
had been submitted for publication 
in Physics Letters B, editor Hendrik Weerts,
in May 2003 with the title
''Neutron lifetime measurements with UCN gravitational trap 
with Be coating''. The authors were
A.\ Serebrov (PNPI and PSI), V.\ Varlamov (PNPI), P.\ Geltenbort (ILL), 
A.\ Kharitonov (PNPI), R.\ Taldaev (PNPI), I.\ Krasnoschekova (PNPI), 
A.\ Vassiljev (PNPI), A.\ Fomin (PNPI), O.\ Zherebtsov (PNPI), 
M.\ Daum (PSI), R.\ Henneck (PSI), K.\ Kirch (PSI), J.\ Butterworth (ILL),
K.\ Schreckenbach (TU Munich). This action was undertaken by the corresponding author, 
A.\,Serebrov, without the agreement of the whole collaboration. There were
major concerns about details of the analysis as they were also pointed out later
by the referee.

The referee's report was obtained from the editor 
on July 18, 2003, where several items should be addressed by modifying
the manuscript. Instead, part of the data from this experiment appear in
Ref.\cite{serebrov}, however, with a surprisingly different and incomplete authorship.

\noindent
{\bf  2) Data in Fig.\,4 of PLA 335 (2005) 327}\\
Data shown in Fig.\,4 originate from joint experiments of a collaboration
between PNPI Gatchina, 
PSI Villigen, ILL Grenoble and 
Phys.\ Dep.\ TU Munich, see ILL proposals 3-14-138 and
3-14-148 above. Results
were submitted for publication in Nucl.\ Instr.\ and Meth.\
in Phys.\ Res.\ A, Manuscript Number RK2004-0124, in December 2004:
`Magnetron-sputtered Be coatings as reflectors for ultracold
neutrons'\cite{genneck}. The list of authors is  
T.\ Brys (PSI), M.\ Daum (PSI), P.\ Fierlinger (PSI), 
A.\ Fomin (PNPI), P.\ Geltenbort (ILL),
R.\ Henneck (PSI), K.\ Kirch (PSI), A.\ Kharitonov (PNPI), 
I.\ Krasnoshekova (PNPI), M.\ Kuzniak (PSI),
M.\ Lasakov (PNPI), A.\ Pichlmaier (PSI), F.\ Raimondi (PSI), 
R.\ Schelldorfer (PSI), A.\ Serebrov (PSI and PNPI),
E.\ Siber (PNPI), R.\ Tal'daev (PNPI), V.\ Varlamov (PNPI), 
A.\ Vasiliev (PNPI), J.\ Wambach (PSI),
O.\ Zherebtsov (PNPI). 

Please compare Fig.\,11 of Ref.\cite{genneck} with Fig.\,4
of Ref.\cite{serebrov}.

\noindent
{\bf Conclusion}\\
The text in Ref.\cite{serebrov} suggests that the mentioned data originate 
from the research of the Ref.\cite{serebrov} authors only (Fig.\,2) or 
from a new experiment with the authors of Ref.\cite{serebrov}, 
see page 4 of Ref.\cite{serebrov}:
''In a new experiment the transmission for fifteen different 
samples was measured for UCN in the energy range ...''.
Various data presented in 
Ref.\cite{serebrov} do not originate from these authors alone. 
They originate, however, from experiments with 
a remarkably different list of participants and contributors. 
Several of these participants learned about
this process after publication only. We do not accept such 
a procedure and state our vehement protest.

Concerning the physics content, the basic approach of Ref.\,\cite{serebrov} has been demonstrated to be
wrong\cite{barabanov}. Furthermore, it was shown experimentally that at least part of the 
``anomalous losses'' can be explained by (i) holes and cracks in the coating surfaces,
(ii) microscopic (or nanoscopic) particles sitting on the surface (dust particles), or
(iii) hydrogen contamination on and in the surface\cite{genneck,fierlinger,depol_1,depol_2,
katerina}.


\begin{thebibliography}{99}
\bibitem{serebrov} A.\,Serebrov et al., Phys.\ Lett.\ A 335 (2005) 327.
\bibitem{PSI}A.\,Kharitonov et al., PSI {\bf $\cdot$} Scientific Report 2002/Volume I, (2003) 29.
\bibitem{genneck} T.\,Bry\'s et al., Nucl.\ Instr.\ and Meth.\ in Phys.\ Res.\ A 551 (2005) 429.
\bibitem{barabanov} A.\,L.\ Barabanov and K.\,V.\ Protasov, Phys.\ Lett.\ A, in press (2005).
\bibitem{fierlinger} P.\,Fierlinger, PhD thesis, University Z\"urich, Switzerland (2005).
\bibitem{depol_1} F.\,Atchison et al., Phys.\ Lett.\ B 625 (2005) 19.
\bibitem{depol_2} F.\,Atchison et al., Nucl.\ Instr.\ and Meth. in Phys.\ Res.\ A 550 (2005) 637.
\bibitem{katerina} E.\,Korobkina et al., Phys.\ Rev.\ B 70 (2004) 35409.

\end{thebibliography}
\end{document}